\documentclass[10pt]{revtex4}               
\usepackage{amsmath,amssymb}
\usepackage{latexsym}
\usepackage{epsfig}

\begin{document}
\title{Black hole thermodynamics in Sharma-Mittal generalized entropy formalism}
\author{S. Ghaffari$^{1}$\footnote{sh.ghaffari@riaam.ac.ir}, A. H. Ziaie$^{1}$\footnote{ah.ziaie@riaam.ac.ir}, H. Moradpour$^1$\footnote{hn.moradpour@gmail.com}, F. Asghariyan$^1$\footnote{fereshteh.asghariyan@yahoo.com}, F. Feleppa$^2$\footnote{feleppa.fabiano@gmail.com}, M. Tavayef$^{1}$\footnote{matin.tavayef@gmail.com}}
\address{$^1$ Research Institute for Astronomy and Astrophysics of Maragha (RIAAM), P. O. Box 55134-441, Maragha, Iran\\ $^2$ Department of Physics, University of Trieste, via Valerio 2, 34127 –Trieste, Italy}

\begin{abstract}
Using the Sharma-Mittal entropy, we study some properties of the
Schwarzschild and Schwarzschild-de Sitter black holes. The results
are compared with those obtained by attributing the Bekenstein
entropy bound to the mentioned black holes. Our main results show
that while the Schwarzschild black hole is always stable in the
micro-canonical ensemble, it can be stable in the canonical
ensemble if its mass is bigger than the mass of the coldest
Schwarzschild black hole. A semi-classical analysis has also been
used to find an approximate relation between the entropy free
parameters. Throughout the paper, we use units $c=G=\hbar=k_B=1$,
where $k_B$ denotes the Boltzmann constant.
\end{abstract}

\maketitle

\section{Introduction}
The Jacobson's pioneering work \cite{J1} reveals the deep
connections between gravity, spacetime and thermodynamics. It says
that the system information (or equally the system entropy),
combined with the thermodynamic laws, can address the
gravitational field equations. The Bekenstein entropy bound is one
of its argument pillars which also plays a crucial role in
investigating the black hole thermodynamics
\cite{J1,Stability1,Stability2}. This entropy is non-extensive, a
property expected by considering the long-range nature of the
gravity \cite{Ts1,fon,non0,abe,nn1,nn2,SME1,5}. Indeed, the
non-extensivity is also brought up in studying the thermodynamic
properties of systems whenever the effects of the size of
reservoir is considered \cite{motivcannon1,motivcannon2}.

The above argument motivates physicists to study the cosmological
and gravitational phenomena in the generalized statistical
formalisms, \textbf{such as non-extensive R\'{e}nyi and Tsallis
entropies}, in which the ordinary probability distribution is
replaced with a power-law distribution leading to generalized
entropies
\cite{non4,non5,non6,non13,non7,EPJC,prd,non8,filipemena,SME2,Tavayef,non21,epjcr,epl,Biro1,thermo1,thermo2,motivcannon3,EPJC1,T1,SME3,SME4,SME5,SME6,pdu}.
\textbf{It has been indicated that using the power-law
distributions of probability is compatible with experimentally
observed power-law tailed particle spectra~\cite{T2,T3,T4,T5}.}
Recently, using the R\'{e}nyi entropy, as a single parameter
generalized entropy \cite{non0,SME1}, the thermodynamic properties
of some black holes has been studied
\cite{Biro1,thermo1,thermo2,motivcannon3,EPJC1}. \textbf{It has
been shown that the Schwarzschild black holes based on the
R\'{e}nyi formalism, in contrast to Boltzmann-Gibbs statistical
mechanics, can be in stable equilibrium in the canonical
ensemble~\cite{EPJC1,motivcannon3}. In spite of the successes of
the non-extensive statistical mechanics, the non-extensive Tsallis
entropy can not satisfy the zeroth law of thermodynamics without
neglecting energy corrections~\cite{T6}}.

On the other hand, its generalization, called the Sharma-Mittal
entropy (SM) \cite{SME1}, \textbf{which is a combination of the
R\'{e}nyi and Tsallis entropies}, leads to interesting results in
the cosmological setup. {\bf Such a kind of generalization} helps
us to describe the current accelerated universe by using the
vacuum energy in a suitable manner
\cite{SME2,SME3,SME4,SME5,SME6}. \textbf{Although, non-extensive
entropies have been used to investigate the thermodynamic
properties of black holes, but none of them consider the
Sharma-Mittal entropy for it.} All of these motivate us to study
the thermodynamic properties of black holes, as strongly coupled
gravitational systems, by employing the SM entropy written as
\cite{SME1,SME2}

\begin{eqnarray}\label{S1}
S_{SM}=\frac{1}{R}\Big((1+\delta S_T)^{\frac{R}{\delta}}-1\Big).
\end{eqnarray}

\noindent Here, $S_T=\frac{A}{4}$, where $A=4\pi r^2$ and $r$
denotes the horizon radius, is the Tsallis entropy \cite{5,SME2}.
Moreover, $R$ and $\delta$ are free unknown parameters {\bf to be} determined by
fitting the results with observations \cite{SME1,SME2}.
\textbf{One can see that for $ R\rightarrow 0 $ and $ R\rightarrow \delta $, the R\'{e}nyi
and Tsallis entropies are recovered, respectively.}

\textbf{It is worth mentioning that, there are two approaches to
demonstrate the thermodynamic properties of black holes in the
non-extensive statistical mechanics. In the first approach,
similar to the standard picture of black hole thermodynamics, the
temperature of a black hole is considered as the Hawking
temperature and so the energy of the system should be generalized.
In another approach, the standard energy of the system is assumed
and a corresponding generalized temperature associated with the
generalized entropy can be obtained. In the present work, we use
the second approach and find the generalized temperature
associated with the Sharma-Mittal entropy to describe the
thermodynamic properties of the Schwarzschild black hole.}

In this paper, we are mainly {\bf interested} to study some thermodynamic
properties of the Schwarzschild (Sch) black holes meeting the SM
entropy bound instead of the Bekenstein entropy bound. After
addressing some general remarks of the SM entropy of black holes
in the next section, we study the temperature and decay time of
the Sch and Schwarzschild-de Sitter (SdS) black holes in the third
section. A semi-classical approach is used, {\bf in the forth section}, to find an approximate
relation between $\delta$ and $R$ parameters of Sch black
holes. Employing the Poincar\'{e} turning
point method \cite{Poincare,Stability1,Stability2,motivcannon3},
the stability of Sch black holes is also studied in both the
micro-canonical and canonical ensembles in Sec.~(\textmd{V}) where
the heat capacity has also been investigated. The last section is
also devoted to a summary. Throughout the paper, we also compare
our results with those obtained by using the Bekenstein entropy.
\section{SM entropy for black hole horizon}
Consider a spherically symmetric static metric

\begin{equation}
ds^2=f(\tilde{r})dt^2-\frac{d\tilde{r}^2}{f(\tilde{r})}-\tilde{r}^2d\Omega^2,
\end{equation}

\noindent where $ t $ and $ \tilde{r} $ are the time and radius
coordinates, respectively, and {\bf $d\Omega^2=d\theta^2+\sin^2\theta d\phi^2$ is the standard line element on a unit two-sphere}. The Unruh temperature corresponding
to the above metric horizon with radius $r$ obtained by solving
$f(r)=0$ equation, is given as~\cite{Unruh}

\begin{equation}
T=\frac{1}{4\pi}f^{\prime}(r).
\end{equation}

For a black hole with mass $M$ (the mass confined by radius $r$), one can use the Clausius formula to reach the entropy
content ($S$) of the black hole (with radius $r$) as \cite{J1}

\begin{eqnarray}
S=4\pi\int_{0}^{M}\frac{dM}{f^{\prime}(r)}=4\pi\int_{0}^{r}\int_{0}^{M}
\delta\Big(f(r,M)\Big)drdM,
\end{eqnarray}

\noindent where we used the fact that $f^{\prime}(r)$ is a
Jacobian for the Dirac-delta constraint \cite{Biro1} to write the
last equality. In fact, the last equality addresses us a
micro-canonical shell in a phase space whose variables are $r$
and $M(\equiv E)$ \cite{Biro1}. For the Bekenstein-Hawking entropy
of the Sch black hole ($S_{BH}$), where $f(r)=1-\frac{2m}{r}$, we
have

\begin{equation}
S_{BH}=\pi r^2,
\end{equation}

\noindent in which $ r=2M $ is the solution of $f(r)=0$.
Therefore, simple calculations lead to \cite{Biro1}

\begin{eqnarray}\label{SBH}
&&S_{BH}=4\pi E^2, ~~~~~ \frac{1}{T_{BH}}=S^{\prime}_{BH}(E)=8\pi E,\nonumber\\
&&C_v=-\frac{{S_{BH}^{\prime}}^2(E)}{S_{BH}^{\prime\prime}(E)}=-8\pi E^2,
\end{eqnarray}

\noindent where prime denotes derivative with respect to $E$.
Moreover, $T_{BH}$ and $C_v$ are Hawking temperature and heat capacity of the black hole at constant volume, respectively
\cite{Biro1}.

\section{Temperature and decay time for Sch and SdS black
holes}

For the Sch black hole, for which $f(r)=1-2M/r=0$, one can use Eqs.~(\ref{SBH}) along with Sharma-Mittal entropy~(\ref{S1}) to obtain

\begin{eqnarray}\label{SE}
&&S_{SM}(E)=\frac{1}{R}\Big((1+4\delta\pi E^2)^{\frac{R}{\delta}}-1\Big),\nonumber\\
&&\frac{1}{T_{SM}}=8\pi E(1+4\delta\pi
E^2)^{\frac{R}{\delta}-1},\nonumber
\end{eqnarray}

\noindent and

\begin{eqnarray}
C_v=-\frac{8\pi E^2(1+4\delta \pi E^2)^\frac{R}{\delta}}{1+8\pi
RE^2-4\delta\pi E^2}.
\end{eqnarray}

\begin{figure}[htp]
\begin{center}
    \includegraphics[width=8cm]{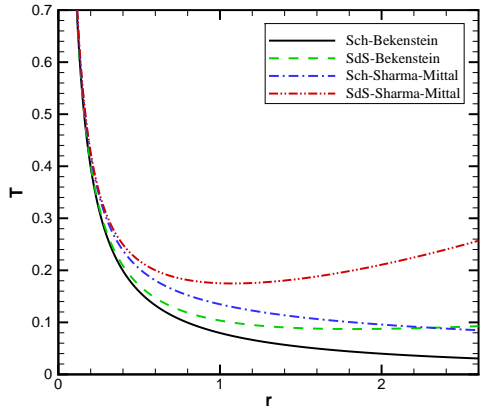}
    \caption{Temperature $T$ versus the black hole radius $r$ for both the Bekenstein and Sharma-Mittal entropies.
    The values of $\lambda=-0.2$, $R=1.2$ and $\delta=1.8$ are adopted.}\label{figT1}
\end{center}
\end{figure}

\noindent Now, using Eq.~(\ref{SE}) one gets

\begin{eqnarray}\label{tem1}
T_{SM}(r)=\frac{1}{4\pi r}\Big(1+\delta\pi
r^2\Big)^{1-\frac{R}{\delta}},
\end{eqnarray}

\noindent as the Hawking temperature. \textbf{In the limiting case
$ R=\delta$, the standard Hawking temperature is recovered. This
limiting case also corresponds to Bekenstein-Hawking entropy which
plays the role of the Tsallis entropy
\cite{EPJC,prd,non8,SME2,epjcr,Biro1,motivcannon3,T6,EPJC1}. This
result implies that the observed deviation from the Hawking
temperature for the Schwarzschild black hole can be attributed to
(or caused by) the non-extensive Sharma-Mittal entropy.} For a SdS
black hole, where $f(r)=1-2M/r+\lambda r^2/2$, in which $\lambda$
denotes the cosmological constant, the $f(r)=0$ equation yields

\begin{equation}\label{M2}
M(r)=\frac{r}{2}\Big(1+\lambda\frac{r^2}{2}\Big),
\end{equation}

\noindent for the mass content of the horizon with radius $r$.
Using the $E\equiv M$ definition together with Eqs. (\ref{tem1}) and (\ref{M2}), the Hawking temperature is evaluated as

\begin{eqnarray}\label{tem2}
T(r)=\frac{\partial M}{\partial
S}=\frac{M^{\prime}(r)}{S^{\prime}(r)}=\frac{1+\frac{3\lambda}{2}r^2}{4\pi
r(1+\delta\pi r^2)^{\frac{R}{\delta}-1}},
\end{eqnarray}

\noindent where prime denotes derivative with respect to radius.
\textbf{In the limit of $R=\delta$, the standard temperature of
Bekenstein-Hawking entropy for the SdS black hole \cite{Biro1} is
regained.} In Fig.~\ref{figT1}, the temperature of Sch and SdS
black holes have been plotted by using the above relations. The
results of using the Bekenstein entropy have also been plotted for
a better comparison.

Considering the Hawking radiation of a black hole as a black body
radiation and bearing the Stefan-Boltzmann law in mind, one has

\begin{eqnarray}
\dot{M}\equiv\frac{dM}{dt}=A\sigma T^4,
\end{eqnarray}

\noindent where $ \sigma $ is the Stefan-Boltzman constant, for
the mass loss rate of a Schwarzschild black hole. It finally leads
to

\begin{equation}
\dot{M}=M^\prime(r)\dot{r}=-4\pi r^2\sigma T^4(r),
\end{equation}

\noindent combined with Eq.~(\ref{tem1}) to obtain the decay time
of the Sch black hole meeting the Sharma-Mittal entropy

\begin{equation}\label{t1}
t_s=\frac{(4\pi)^3}{\sigma}\int_{0}^{r}r^2\Big(1+\delta\pi r^2\Big)^{\frac{4R}{\delta}-4}dr.
\end{equation}

\begin{figure}[htp]
\begin{center}
    \includegraphics[width=8cm]{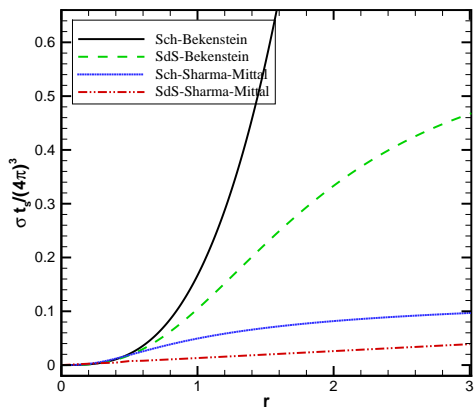}
    \caption{$t_s(r)$ for both the Bekenstein and Sharma-Mittal entropies. We have set $\lambda$=0.2, $R=0.7$ and $\delta=1.5$.}\label{figt1}
\end{center}
\end{figure}

\noindent Hypergeometric functions are the solutions of the above
integral. For the SdS black holes, calculations lead to

\begin{eqnarray}\label{t2}
dt_s=\frac{(4\pi)^3r^2(1+\delta\pi r^2)^{\frac{4R}{\delta}-4}}{\sigma(1+\frac{3}{2}\lambda r^2)^3}dr,
\end{eqnarray}

\noindent for the time decay, which yields again hypergeometric
functions. Fig.~\ref{figt1} includes the $t_s(r)$ of the Sch and
SdS black holes for both the Bekenstein and Sharma-Mital
entropies.

\section{A semi-classical investigation for entropy parameters of Sch black holes}

Consider a Sch black hole whose temperature is the minimum value
of Eq.~(\ref{tem1}) leading to $r_0^2=\frac{1}{\pi(\delta-2R)}$
for its radius. From classical point of view, its energy is
$E_0=M(r_0)=\frac{r_0}{2}$. On the other hand, since this state
has the minimum temperature, one may also approximate it with
quantum mechanical considerations \cite{Biro1}. As a toy model, we
look at this state as the ground state of a harmonic oscillator
with energy $E=\frac{\omega}{2}=\frac{\pi}{\Lambda}$ (in the unit
of $\hbar=c=1$). Here, $\Lambda$ denotes the wavelength of the
system, and $2r_0$ (the black hole diameter) is a suitable
approach for $\Lambda$ \cite{Biro1} yielding
$E\approx\frac{\pi}{2r_0}$. Employing $E_0\approx E$
assumption, and bearing the $r_0^2=\frac{1}{\pi(\delta-2R)}$
relation in mind, we reach

\begin{equation}
\delta\approx\frac{1}{2\pi^2}+2R,
\end{equation}

\noindent as an approximate relation between $\delta$ and $R$ in
the mentioned situation.
\section{Stability of Sch black hole}

\begin{figure}[htp]
\begin{center}
    \includegraphics[width=8cm]{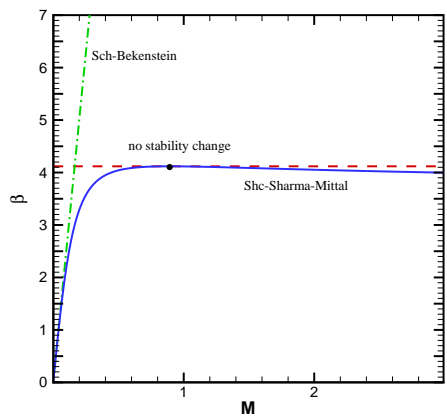}
    \caption{The $\beta(M)$ curves of a Sch black hole within the frameworks of
    Bekenstein and Sharma-Mittal entropies in the microcanonical regime when $R=0.7$ and
    $\delta=1.5$. No vertical tangent occurs in either case.}\label{figs1}
\end{center}
\end{figure}

Here, we are going to study the stability of Sch black holes,
meeting the Sharma-Mittal entropy bound, in both the
micro-canonical \cite{Poincare,Stability1,Stability2,motivcannon3}
and canonical \cite{motivcannon1,motivcannon2,motivcannon3}
approaches.
\textbf{It has been shown that in the Boltzmann-Gibbs picture the Sch black holes
cannot be in stable equilibrium which implies that the canonical ensemble can not
be used when the gravitational interaction is included in the standard statistical mechanics~\cite{Hawking1,Hawking2}.}
In the micro-canonical framework, it has been assumed
that the Sch black hole is isolated, and thus, $\beta$ and $M$ are
the corresponding conjugate thermodynamic variables which have the
mutual relation \cite{Poincare,Stability1,Stability2,motivcannon3}
\begin{equation}
\beta=\frac{\partial S}{\partial E},
\end{equation}

\noindent where $S$ denotes the system entropy. Hence,
$\beta=\frac{1}{T}$ and

\begin{equation}
\beta=8\pi M(1+4\delta\pi M^2)^{\frac{R}{\delta}-1},
\end{equation}

\noindent which recovers the result of employing the Bekenstein
entropy ($\beta=8\pi M$) at the appropriate limit $R=\delta$
\cite{motivcannon3}. {\bf As} long as the slope of the $\beta$ curve
(versus $M$) is not vertical, the configuration under study does
not {\bf admit} any turning point meaning that the system remains stable
\cite{Poincare,Stability1,Stability2,motivcannon3}. In
Fig.~\ref{figs1}, $\beta$ versus $M$ has been plotted for both the
Bekenstein and Sharma-Mittal entropies indicating that the
isolated Sch black hole, \textbf{similarly to the standard Boltzmann-Gibbs picture},
is always stable \textbf{against spherically symmetric perturbations} in both cases
({\bf no} turning point is observed).

\begin{figure}[htp]
\begin{center}
    \includegraphics[width=8cm]{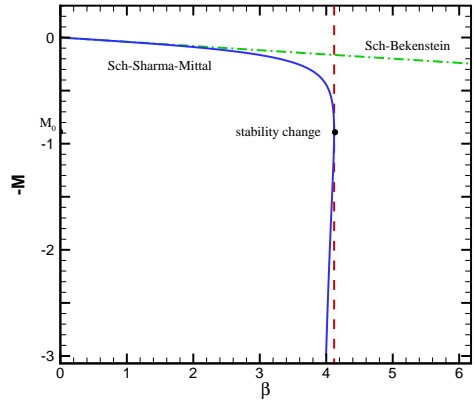}
    \caption{The $-M(\beta)$ curves of a Sch black hole within the Boltzmann
    and Sharma-Mittal frameworks in the canonical ensemble when $R=0.7$ and $\delta=1.5$.}\label{figs2}
\end{center}
\end{figure}

The Helmholtz free energy formalism ($F$) is the backbone of the canonical survey of a black hole in contact with a thermal bath, which is defined as
\cite{motivcannon1,motivcannon2,motivcannon3}

\begin{equation}
F=\frac{S}{\beta}-M,
\end{equation}

\noindent where $\beta$ and $-M(\equiv\frac{d(\beta
F)}{d\beta})$ are the conjugate thermodynamic variables
\cite{motivcannon1,motivcannon2,motivcannon3}. As it is apparent
from Fig.~\ref{figs2}, the tangent of the $-M(\beta)$ curve is
vertical when $M=M(r_0)\equiv M_0$, or equally, the temperature of
the Sch black hole is the minimum value of Eq.~(\ref{tem1})
($T_0=1/\sqrt{\pi(\delta-2R)}$).
\textbf{This means that, Schwarzschild black hole in the Sharma-Mittal formalism,
in contrast to the  Bekenstein-Hawking entropy, can be stable for $ M>M_0 $. We therefore conclude that black holes with $M>M_0$ have positive specific heat capacity and are stable while those with $ M<M_0 $ are unstable.
These results are comparable with those obtained in~\cite{motivcannon3} for the Schwarzschild black holes described by the R\'{e}nyi formalism.\\}

\begin{figure}[htp]
\begin{center}
    \includegraphics[width=8cm]{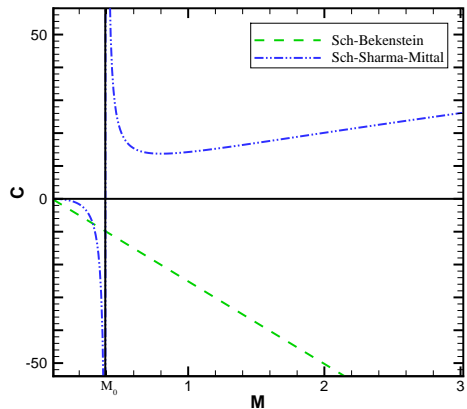}
    \caption{Heat capacity of Sch black hole for both the Bekenstein and Sharma-Mittal entropies. Here, $R=0.6$ and $\delta=1.8$.}\label{figC}
\end{center}
\end{figure}

The sign of the heat capacity $C(\equiv\frac{dM}{dT})$ can also
help us in determining the stability of the black holes (the
Hessian analysis) \cite{motivcannon3}. From Fig.~\ref{figC}, one
can see that, unlike the Bekenstein formalism, the sign of the
heat capacity is changed {\bf at} $M_0$ point whenever the
Sharma-Mittal entropy is attributed to the Sch black hole. Indeed,
for $M>M_0$, the heat capacity is positive and meaningful, while
for $M<M_0$, the heat capacity is negative meaning that such Sch
black holes {\bf violate} {\bf the laws of thermodynamics}, a result obtainable
in the Bekenstein formalism for which the heat capacity of Sch
black hole is always negative.

\section{CONCLUSION}

Our main focus, in this work, was to study some properties of a Sch black hole
meeting the Sharma-Mittal entropy and compare them with those
obtained by employing the Bekenstein entropy. Firstly, we
addressed a time that the Sch and SdS black holes need to be
completely evaporated through the Hawking radiation mechanism.
Considering a Sch black hole with minimum temperature ($T_0$) {\bf together with} using a semi-classical analysis, we could obtain a bound on the
entropy parameters $\delta$ and $R$. It has also been found out
that, in the Sharma-Mittal formalism, a Sch black hole has
positive heat capacity and is stable in the canonical ensemble
framework, if its mass is bigger than $M_0$ which is the mass of a
Sch black hole with temperature $T_0$. \textbf{This result is contrary to the standard Boltzmann-Gibbs statistical mechanics.}
The stability analysis also {\bf shows} that {\bf similar to} the Bekenstein case,
a Sch black hole meeting the Sharma-Mittal entropy bound is always stable
in {\bf the framework of} micro-canonical ensemble.
\acknowledgments{The work of S. Ghaffari has been supported
    financially by Research Institute for Astronomy and Astrophysics of
    Maragha (RIAAM).}

\end{document}